\documentclass{aastex62}

\received{}
\revised{}
\accepted{}
\submitjournal{ApJ}

\shorttitle{Small-scale Turbulence as the Precursor of a Filament Eruption}
\shortauthors{Seki et al.}

\begin{document}

\title{Small-scale Turbulent Motion of the Plasma in a Solar Filament as the Precursor of Eruption}

\correspondingauthor{Daikichi Seki}
\email{seki@kwasan.kyoto-u.ac.jp}

\author[0000-0001-6390-5809]{Daikichi Seki}
\affil{Graduate School of Advanced Integrated Studies in Human Survivability, Kyoto University, Sakyo, Kyoto, 606-8306, Japan}
\affil{Astronomical Observatory, Kyoto University, Yamashina, Kyoto, 607-8471, Japan}
\affil{Department of Applied Mathematics and Theoretical Physics, Wilberforce Road, Cambridge, CB3 0WA, United Kingdom}
\affil{Centre for the Study of Existential Risk, University of Cambridge, 16 Mill Lane, Cambridge, CB2 1SB, United Kingdom}

\author{Kenichi Otsuji}
\affil{Astronomical Observatory, Kyoto University, Yamashina, Kyoto, 607-8471, Japan}
\affil{Applied Electromagnetic Research Institute, National Institute of Information and Communications Technology, Koganei, Tokyo 184-8795, Japan}

\author{Hiroaki Isobe}
\affil{Faculty of Fine Arts, Kyoto City University of Arts, Nishikyo, Kyoto, 610-1197, Japan}

\author{Giulio Del Zanna}
\affil{Department of Applied Mathematics and Theoretical Physics, Wilberforce Road, Cambridge, CB3 0WA, United Kingdom}

\author{Takako T. Ishii}
\affil{Astronomical Observatory, Kyoto University, Yamashina, Kyoto, 607-8471, Japan}

\author{Takahito Sakaue}
\affil{Astronomical Observatory, Kyoto University, Yamashina, Kyoto, 607-8471, Japan}

\author{Kiyoshi Ichimoto}
\affil{Astronomical Observatory, Kyoto University, Yamashina, Kyoto, 607-8471, Japan}

\author{Kazunari Shibata}
\affil{Astronomical Observatory, Kyoto University, Yamashina, Kyoto, 607-8471, Japan}



\begin{abstract}
A filament, a dense cool plasma supported by the magnetic fields in the solar corona, often becomes unstable and erupts.
It is empirically known that the filament often demonstrates some activations such as 
a turbulent motion prior to eruption.
In our previous study \citep{seki17}, we analysed the Doppler velocity of an H$\alpha$ filament
 and found
that the standard deviation of the line-of-sight-velocity (LOSV) distribution in a filament, which indicates the increasing amplitude of the small-scale motions, increased prior to the onset of the eruption.
Here, we present a further analysis on this filament eruption, which initiated approximately at 03:40UT on 2016 November 5 in the vicinity of NOAA AR 12605.
It includes a coronal line observation and the extrapolation of the surrounding magnetic fields.
We found that both the spatially averaged micro-turbulence inside the filament 
and the nearby coronal line emission
increased 6 and 10 hours prior to eruption, respectively.
In this event, we did not find any significant changes 
in the global 
potential-field configuration preceding the eruption for the past 2 days, which indicates that there is a case in which it is difficult to predict the  eruption only by tracking the extrapolated global magnetic fields.
In terms of space weather prediction, our result on the turbulent motions in a filament could be used as the useful precursor of a filament eruption.
\end{abstract}



\section{Introduction} \label{sec:intro}
A filament eruption, a spectacular erupting phenomenon of dense cooler plasma, often initiates on the solar surface.
A filament is a dense (10$^{9}$--10$^{11}$ cm$^{-3}$) and cool (10$^{4}$ K) plasma floating in the solar corona (density $\sim10^{8}$ cm$^{-3}$, and temperature $\sim10^{6}$ K) supported by magnetic fields. 
The plasma is believed to be in equilibrium state due to the balance between gravitational force and Lorentz force.
Several models of the magnetic-field configuration supporting the dense and cool plasma have been proposed (e.g. \cite{tandberg1995nature}).
The so-called Kippenhahn-Schul\"uter (KS) model first proposed by \cite{ksmodel} suggested that the magnetic-field topology has a concave-upward shape where the dense plasma is condensed.
In another model called Kuperus-Raadu (KR) first proposed by \cite{krmodel}, the magnetic field has a helical configuration similar to a flux rope. 
In the KR model, the plasma is condensed at the bottom of the flux rope.

A filament eruption is often preceded by various dynamical motions called filament activations \citep{SmithRamsey64, parenti14, tandberg1995nature, sterling11}.
Slow ascending motion of a filament, typically with a velocity of a few km s$^{-1}$ and a duration time of tens of minutes for active region filament and hours for quiescent filament, has been widely reported in studies prior to the eruption (e.g. \cite{ohshi97,slingmo04}).
A turbulent motion was also reported prior to a filament eruption \citep{tandberg1995nature} and, more generally, a solar flare.
\cite{2001ApJ...549L.245H} observed that non-thermal velocity inside an active region corona increased prior to the flare on 1993 October 3 and suggested that this is the indicator of turbulent changes in the active region.

In our previous study \citep{seki17}, we reported that the standard deviation of line-of-sight (LOS) velocity distribution of small-scale plasma motions inside a filament increased as the filament was reaching the eruption around 03:40 UT on 2016 November 5.
By using the Solar Dynamics Doppler Imager (SDDI)  \citep{2017SoPh..292...63I} installed on the Solar Magnetic Activity Research Telescope (SMART) \citep{2004SPIE.5492..958U} at the Hida Observatory, we monitored the Doppler velocity map of the filament from 29 hours before the onset of the eruption.
As a result, we determined that the standard deviation increased to 3--4 km s$^{-1}$ 6 hours prior to the eruption, whilst it was 2--3 km s$^{-1}$ from 29 to 21 hours prior to the eruption.
The average LOS velocity was approximately constant around 0 km s$^{-1}$.
Thus, we concluded that this broadening LOS velocity distribution (increase of the standard deviation) 
could reflect information on the preceding turbulent plasma motion inside a filament.

In this paper, we present further detailed analysis on this event including the extrapolation of the surrounding 
potential
fields and the estimate of the intensity in coronal emission lines.
\cite{2006PhRvL..96y5002K} investigated the ideal magnetohydrodynamic (MHD) instability called the torus instability first discussed in \cite{bate78} in the situation of low-beta magnetised plasma.
They found that critical conditions to favour the instability were related to the so-called ``decay index'', which is defined as the minus gradient of the unsigned horizontal component of the overlying magnetic field with respect to height.
In this study, we investigated the decay index of potential fields surrounding the filament to see whether the torus instability was favourable to be initiated.

In addition, we performed an observational analysis 
using the SDO/AIA 94, 211, 171, 193 and 304 bands to investigate the variation of intensity in the coronal emission lines.
The destabilisation of a flux rope containing a filament by magnetic reconnections plays a key role in initiation of its eruption \citep{1995JGR...100.3355F,2000ApJ...545..524C}.
If small-scale magnetic reconnections take place prior to the eruption, rising intensity in coronal (high-temperature) emission lines will be observable.
Those SDO/AIA bands enable us to estimate high-temperature emissions such as Fe \textrm{XIV} (a few 10$^{6}$ K) and Fe \textrm{XVIII} ($\sim7$ MK) \citep{dz13}.

This paper is organised as follows: 
In Section \ref{sec:obs}, we describe observations and methods used for our analysis with a brief overview of this event. 
In Section \ref{sec:res}, we present the results of our decay-index investigation and observational analysis 
in multiple wavelengths in SDO/AIA bands.
In Section \ref{sec:disc}, we deliver some interpretations of the turbulent motions and the trigger and evolution of the eruption and the possibility of the application of our results to space weather prediction.

\section{Observations and Methods} \label{sec:obs}
\subsection{Data}
In this study, we used a multiple-wavelengths observation of the Sun by the Atmospheric Imaging Assembly \citep{2012SoPh..275...17L} onboard the Solar Dynamics Observatory (SDO).
The SDO/AIA captures the full-disk Sun routinely in 10 wavelengths including 
94, 211, 171, 193 and 304 \AA\ 
with a time cadence of 12 seconds and a spatial sampling of 0.6 arcsec per pixel. 
 
LOS magnetograms of the Sun taken by the Helioseismic and Magnetic Imager (HMI) onboard the SDO \citep{scherrer12} and a synoptic chart of the photospheric radial magnetic fields inputted by HMI LOS magnetograms were also used.
The SDO/HMI observes the solar full-disk LOS magnetogram with a time cadence of 45 seconds and a spatial sampling of 0.5 arcsec per pixel.
Synoptic HMI charts are remapped ``radial'' component of magnetograms onto the Carrington coordinate grid. 
Here, ``radial'' means that the observed HMI LOS magnetograms are assumed to be the LOS component of purely radial magnetic fields.
Thus, for the computation of this imputed radial magnetic-field component, the HMI LOS magnetograms are divided by the cosine of the angle from the disk centre.

The full-disk images in H$\alpha$ line centre and its wings captured by the SDDI
were used to see the morphology of the target filament and to compute its LOS velocity. 
The SDDI installed on the SMART \citep{2017SoPh..292...63I} at the Hida Observatory, Kyoto University, has been conducting routine observations since 2016 May 1.
It captures the solar full-disk images in 73 wavelengths with a step of 0.25 \AA\ from H$\alpha$ line centre - 9.0 \AA\ to H$\alpha$ line centre + 9.0 \AA\, i.e., at 36 positions in the blue wing, H$\alpha$ line centre, and 36 positions in the H$\alpha$ red wing. 
Each image is obtained with a time cadence of 15 seconds and a spatial sampling of 1.23 arcsec per pixel. 
A part of daily observational data (H$\alpha$ centre, $\pm$ 2.0, $\pm$ 1.25, $\pm$ 0.5, and +3.5 \AA) is always available on the website (https://www.hida.kyoto-u.ac.jp/SMART/T1.html) from 2016 May 1 to the present
\footnote{
The full version of the data (73 wavelengths) is also available.
Please contact us (data\_info@kwasan.kyoto-u.ac.jp), if you are interested in using it.
}.

\subsection{Overview}
The observed filament first appeared from the solar east limb on 2016 October 28 as a prominence.
Its highest spine was measured as 33 Mm above the limb.
It was located at the latitude of N27 degree and in the vicinity of NOAA AR 12605 (see Figure \ref{di_height}), and was tilted by an angle of approximately 45 degree with respect to the south-north direction.
Its length was approximately 112 Mm measured in H$\alpha$ centre.

The filament started to erupt around 03:00 UT to the north on November 5 and totally disappeared in H$\alpha$ line centre around 03:40 UT.
Two-ribbon brightening was observed in both H$\alpha$ line centre and the SDO/AIA 304 in the vicinity of the filament location after eruption 
(see Figure \ref{fe14} and \ref{shear}).
This filament eruption was associated with a B1.1-class flare in the GOES soft X-ray, which peaked at 04:30 UT 
(see Figure 2 in \cite{seki17}).

A CME was observed at 04:36 UT on November 5 in the SOHO/LASCO C2 \citep{lasco} with a linear speed of 403 km s$^{-1}$.
A moderate geomagnetic disturbance on November 10--11, which peaked at -59 nT in Dst \citep{dstindex} at 18:00 UT on November 10, was also observed.
According to Richardson \& Cane Catalogue \citep{cr03}, an interplanetary CME shock was first detected on 2016 November 9 at 06:04UT, which is consistent with the estimated arrival time of the CME (2016 November 9 11:44UT) derived from the simple empirical model provided by \cite{gopal00}.
After the arrival of the shock, the interplanetary CME plasma and the magnetic fields were observed by the Faraday Cup instrument of the Solar Wind Experiment onboard the Wind spacecraft\citep{02kasper} from 2016 November 10 00:00UT to 2016 November 10 16:00UT on the basis of the occurrence of abnormally low proton temperature and the reduced fluctuations and organisation in the interplanetary magnetic fields \citep{cr03}.
Thus, this geomagnetic disturbance can be attributed to the CME associated with the filament eruption.
\subsection{Extrapolation of magnetic fields}
To investigate the decay index surrounding the filament, we computed potential fields from 
the radial magnetic-field components of the photosphere. 
The decay index calculated from potential fields is commonly used for the stability analysis (e.g. \cite{filippov13, joshi14a, joshi14b, li16}).
In this study, we set the critical decay index as one, above which the torus instability is considered to be more favourable, following previous observational studies \citep{filippov13, joshi14a}.

It should be noted that it was necessary to compute potential fields for the whole Sun unlike other cases of decay-index analysis in which potential fields are extrapolated only for a subregion of the Sun (e.g. \cite{filippov13}), because the volume to be considered
 is too large to ignore the sphericity of the Sun.
Moreover, for the objective of tracking the temporal evolution of the global magnetic fields with a time cadence of 1 day, a synoptic chart is not suitable for the boundary condition of potential-field-source-surface (PFSS) extrapolation, because it is provided once in $\sim$ 27 days (rotation period of the Sun).
Therefore, assuming that the unobservable hemisphere was equivalent to the HMI synoptic map, we used ``patched HMI synoptic charts''
 (described later)
 as a boundary condition for the PFSS extrapolation.

Here, we describe how to construct a boundary image used for extrapolation.
There are three steps; first, we inputted the radial component of magnetic fields from HMI LOS magnetogram, assuming that HMI measures the line-of-sight component of a purely radial magnetic field.
Secondly, we extracted a region of a magnetogram between heliocentric latitudes and longitudes of $\pm$ 60 degrees.
Thirdly, we patched the region in the form of a heliocentric spherical coordinate to the HMI synoptic chart for Carrington Rotation 2183.

We used Potential Field Source Surface Solver provided by Yeates, A. R. \citep{yeates17} to extrapolate potential fields \citep{vB00}.
This Python-based code solves the basic equations for magnetic fields by using a finite-difference method with an assumption that the electric currents are negligible in a spherical shell.
For more details, see \url{https://github.com/antyeates1983/pfss}.
We chose 2.0 solar radius above the photosphere and 7 Mm as the height of source surface and the grid size of radius (height), respectively.
Grid sizes of zenith and azimuth angle were taken to be 0.5 degree.

\subsection{Cloud model}
To compute the LOS velocity and the micro-turbulence inside a filament, we utilised a cloud model first proposed by \cite{1964beckers}.
Assuming that (1) the source function is constant along the wavelengths and (2) along the LOS direction and that (3) the line absorption coefficient is a Gaussian shape, this model enables us to determine four physical parameters of the plasma cloud, the source function, the Doppler width, the Doppler shift, and the optical depth \citep{2003PASJ...55..503M, 2003PASJ...55.1141M, 2010PASJ...62..939M, 2017ApJ...836...33C, sakaue18, seki17, seki19}.
The LOS velocity ($v_{los}$) and the micro-turbulence ($\xi$) at each pixel were calculated from the Doppler shift ($\Delta\lambda_{S}$) and the Doppler width ($\Delta\lambda_{D}$), respectively, based on the equations below \citep{2003PASJ...55..503M};
\begin{eqnarray}
v_{los} = c \frac{\Delta\lambda_{S}}{\lambda_{0}},\\ 
\Delta\lambda_{D} = \frac{\lambda_{0}}{c} \sqrt{\xi^{2} + \frac{2k_{B}T}{m_{p}}},
\end{eqnarray}
where $c, \lambda_{0}, k_{B}, T$, and $m_{p}$ are the velocity of light, the wavelength of H$\alpha$ line centre (6562.808 \AA), Boltzmann constant,  a fixed temperature of 10$^{4}$ K in the filament, and the mass of proton, respectively.
(For further information, refer to our previous works \citep{seki17, seki19}.)

\section{Result}\label{sec:res}
\subsection{Decay index}
\begin{figure}[ht!]
\gridline{
	\fig{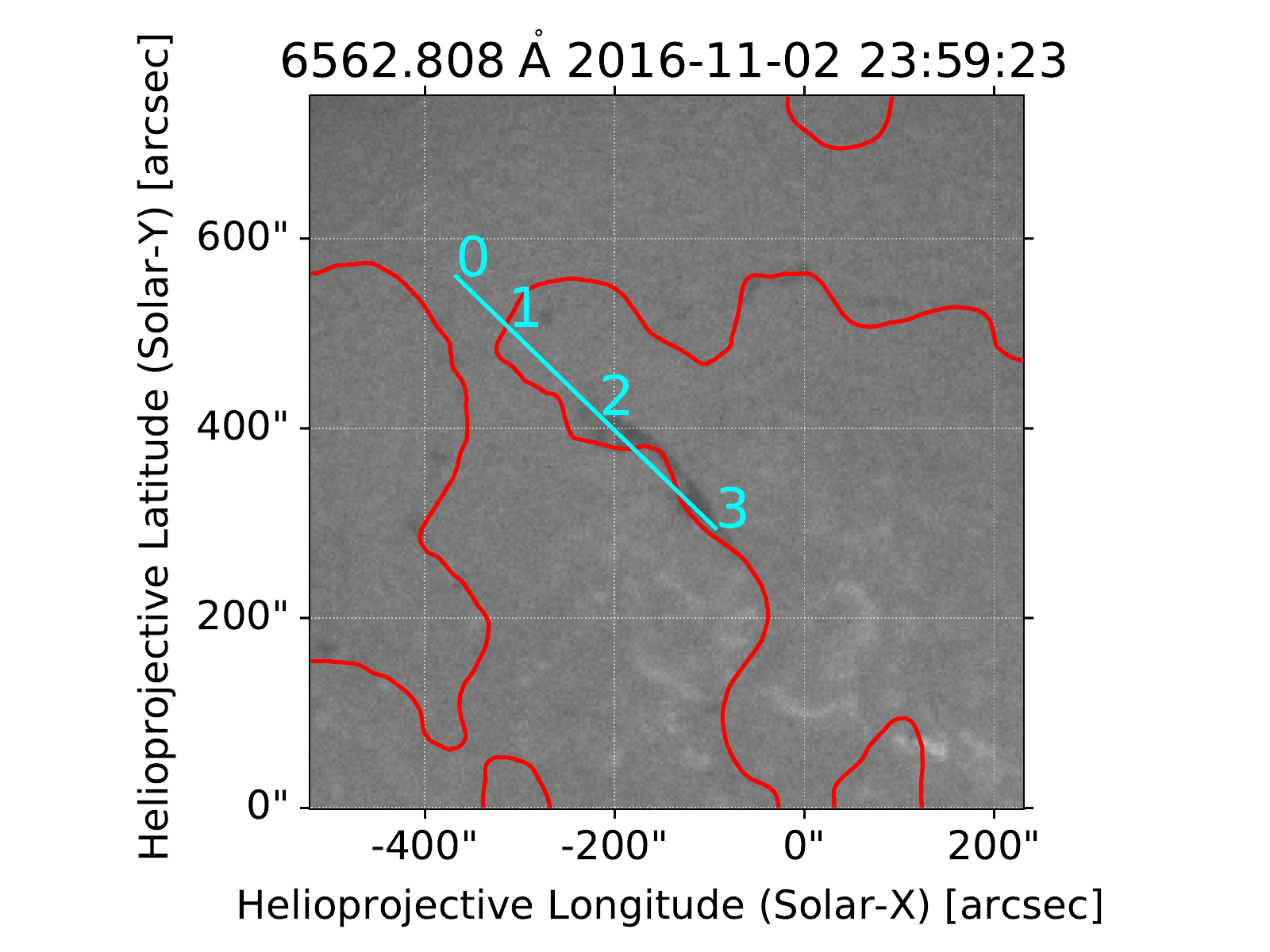}{0.45\textwidth}{}
	\fig{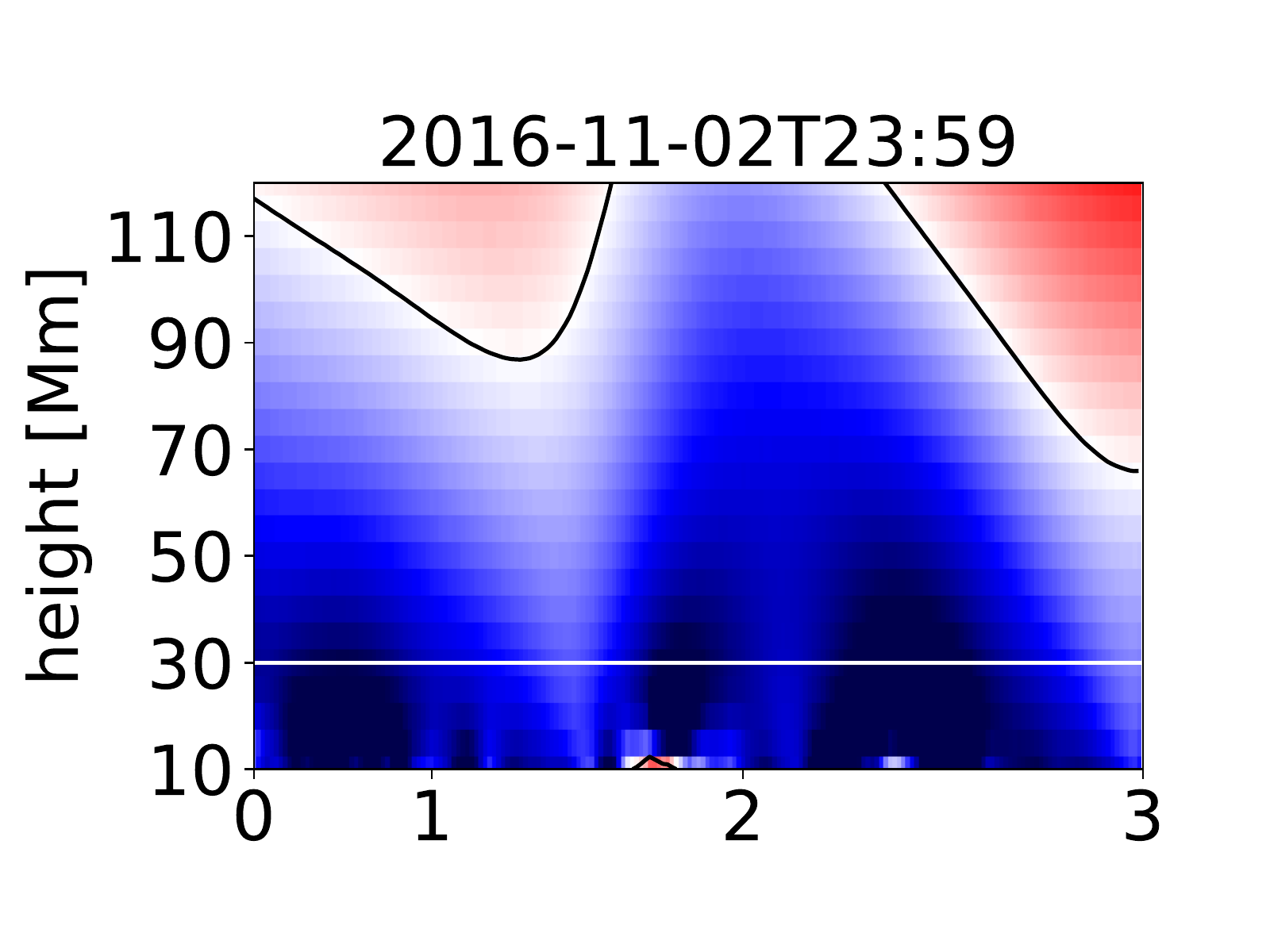}{0.45\textwidth}{}
	}
\gridline{
	\fig{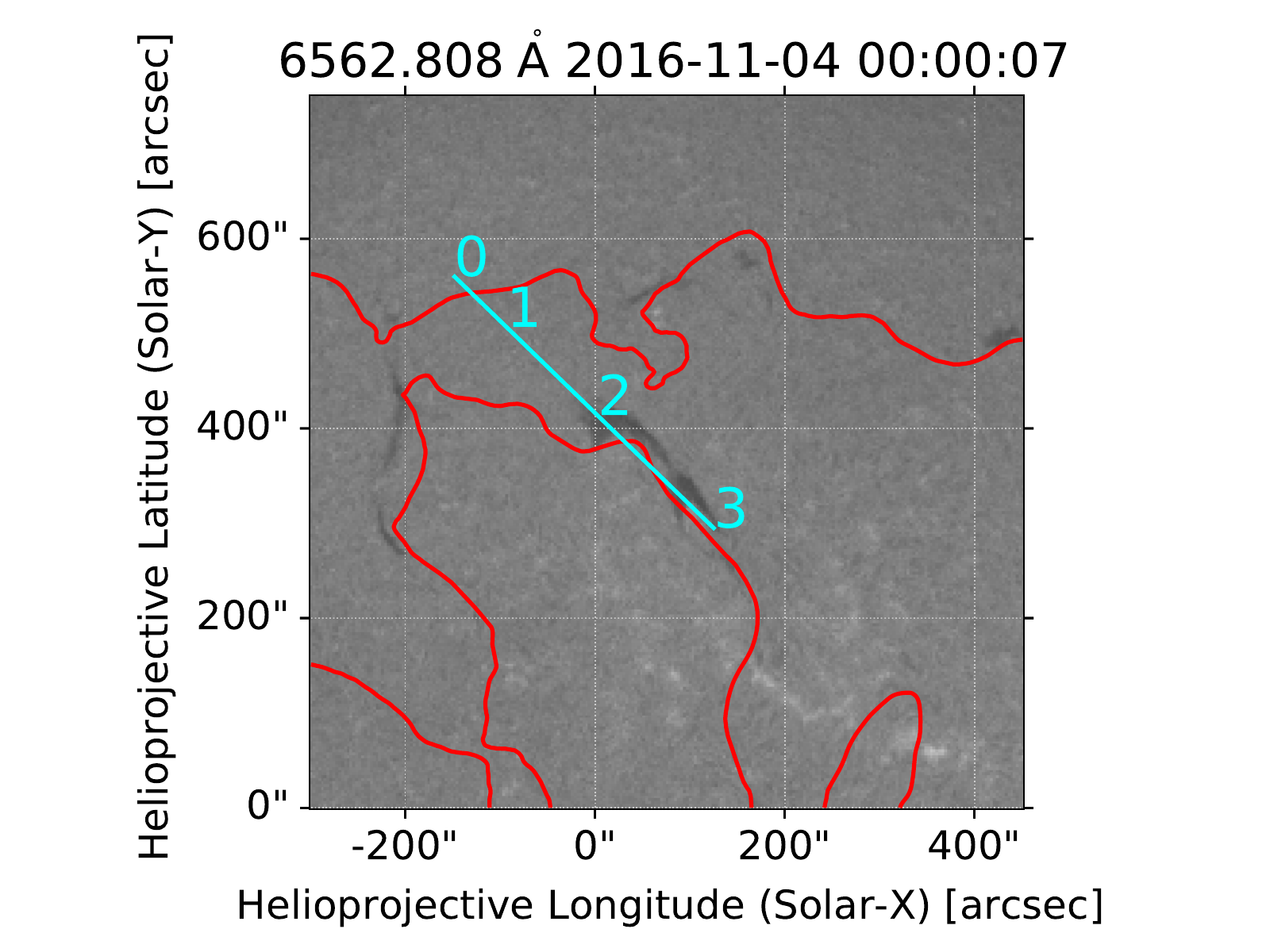}{0.45\textwidth}{}
	\fig{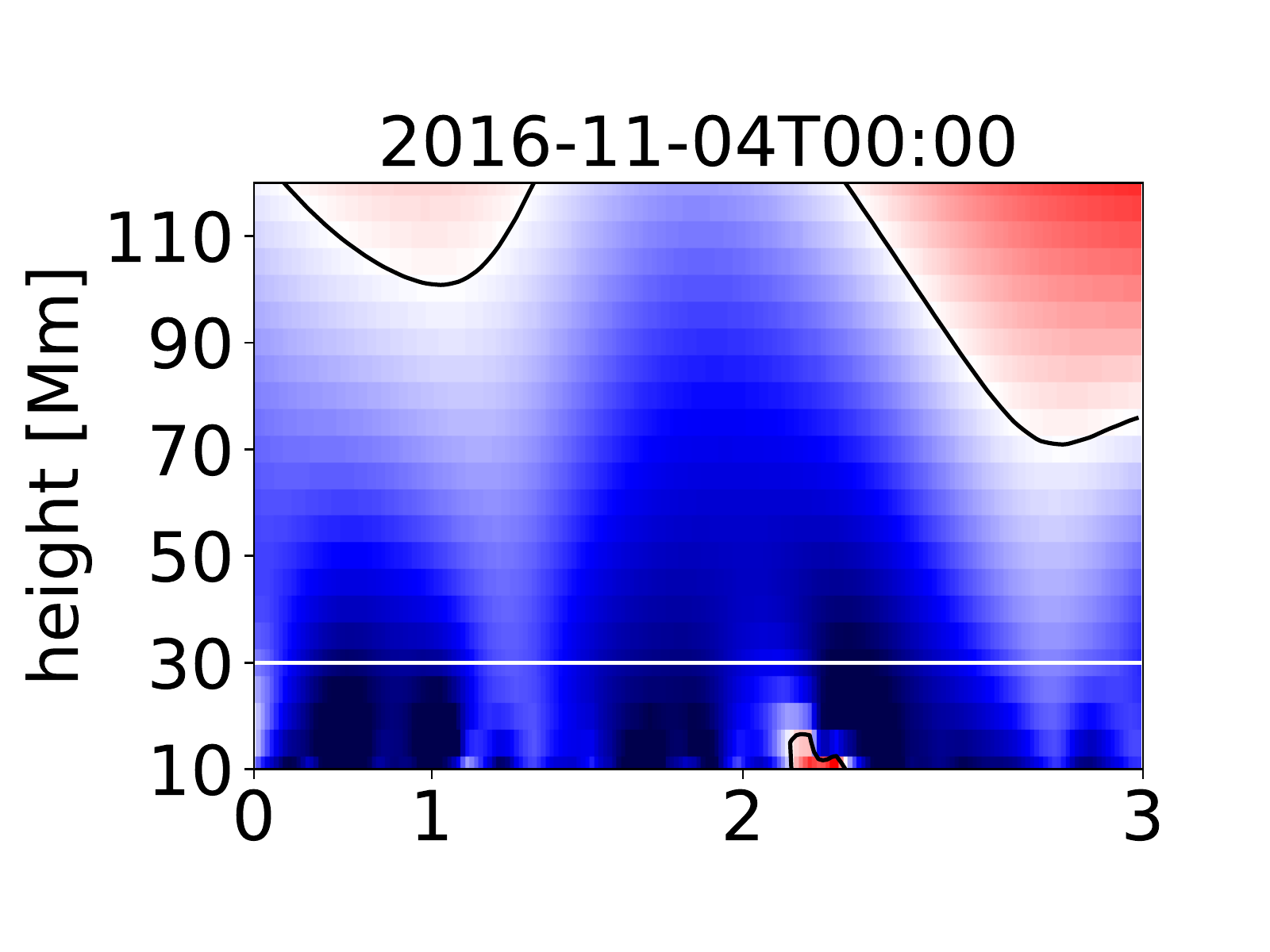}{0.45\textwidth}{}
	}
\gridline{
	\fig{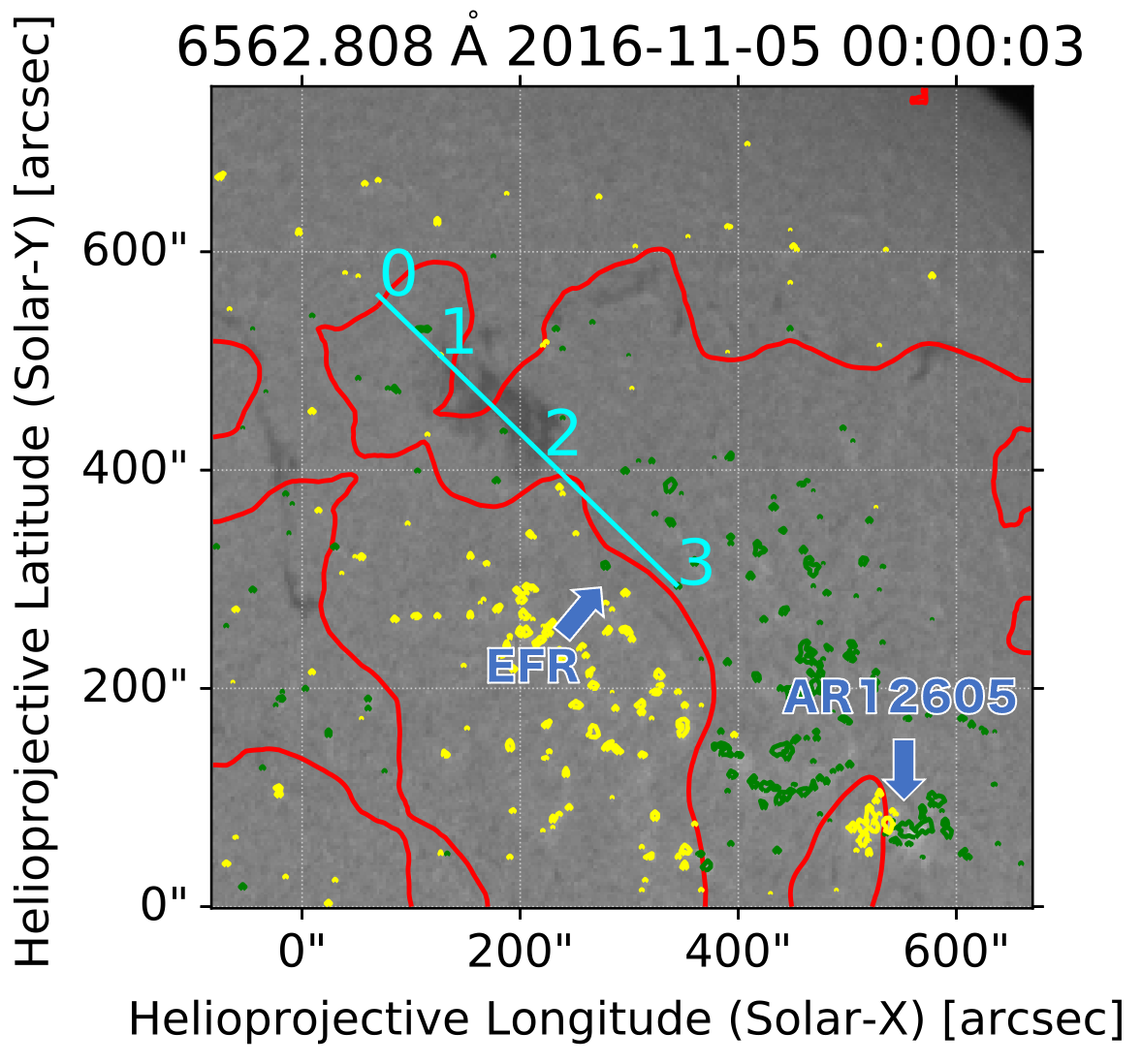}{0.4\textwidth}{}
	\fig{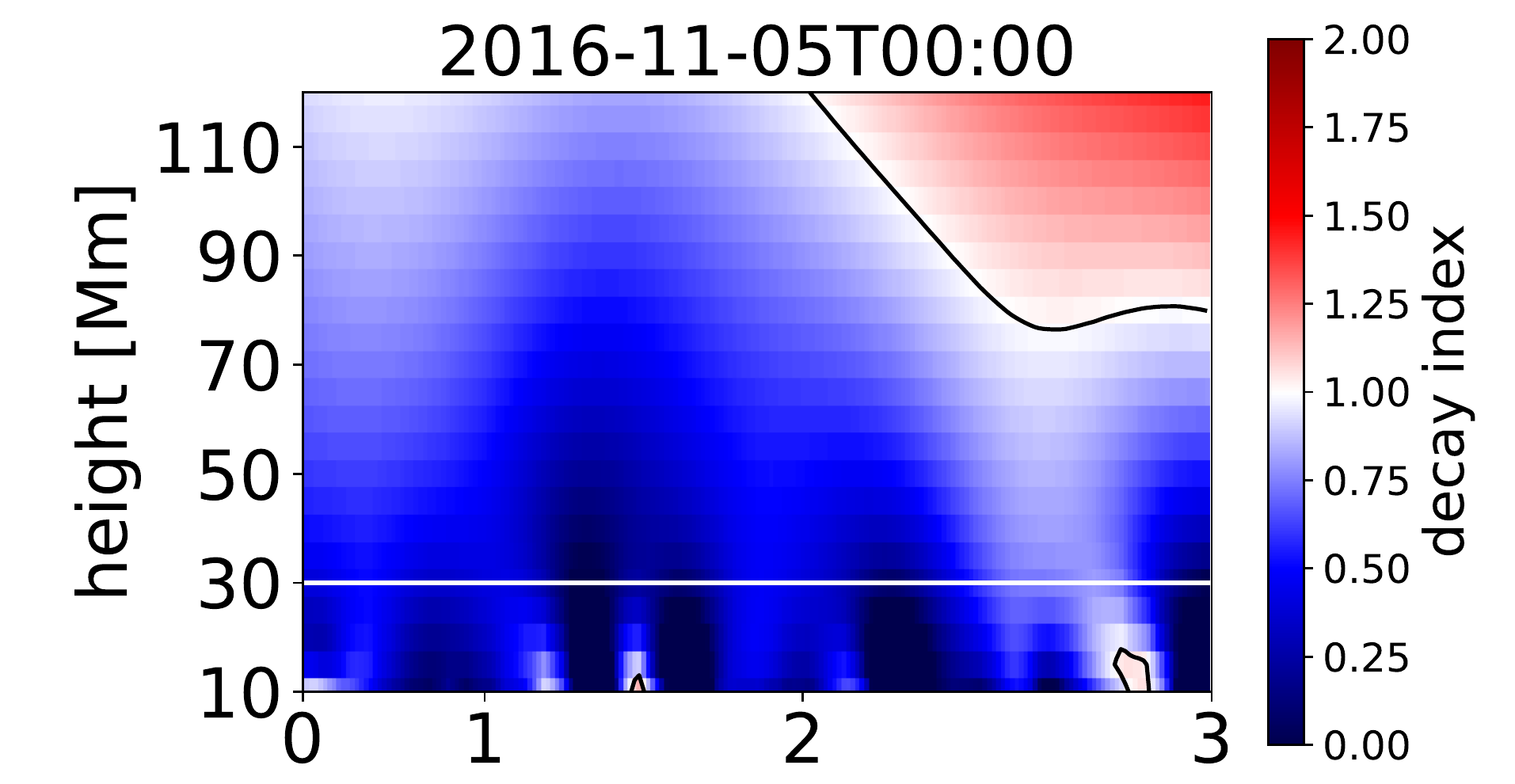}{0.6\textwidth}{}
	}
\caption{
Left: H$\alpha$ centre images observed by SDDI at three different times superimposed by photospheric PIL (red line).
Cyan line in each panel corresponds to the cross section of the right panel.
Yellow and green contours indicate the HMI LOS magnetogram at $\pm$100 G.
For EFR, see Figure \ref{efr}.
Right: Side view of three-dimensional distribution of decay index. 
Each digit at the bottom axis corresponds to the digit on the cyan line of the left panel.
White and black lines corresponds to the approximate height of the filament (30 Mm) and the contour where decay index is 1, respectively.
}
\label{di_height}
\end{figure}

Figure \ref{di_height} demonstrates the decay-index distribution in the vicinity of the filament on November 3, 4 and 5 at 00:00 UT.
Left panels show H$\alpha$ images superimposed by the photospheric polarity inversion lines (PILs) denoted by red lines and the cross sections of the right panels depicted as cyan lines.
Each right panel exhibits the side view of the three-dimensional decay index along the cyan line in the left panel.
From top to bottom, we can recognise a similar decay-index distribution with time.
Generally, the decay index had larger values in the higher location.
In the vicinity of the filament, the decay index was always below one.
Note that the filament in H$\alpha$ laid between the positions 2 and 3 on November 3 and 4, whilst it was located between 1 and 2 on November 5.
The emerging flux region (EFR; see Figure \ref{efr}) appeared in the vicinity of the position 3.
The height of the filament was assumed to be 30 Mm because its highest spine was measured to be 33 Mm above the limb as a prominence on 2016 October 28.

\subsection{Micro-turbulence}
\begin{figure}[ht!]
\plotone{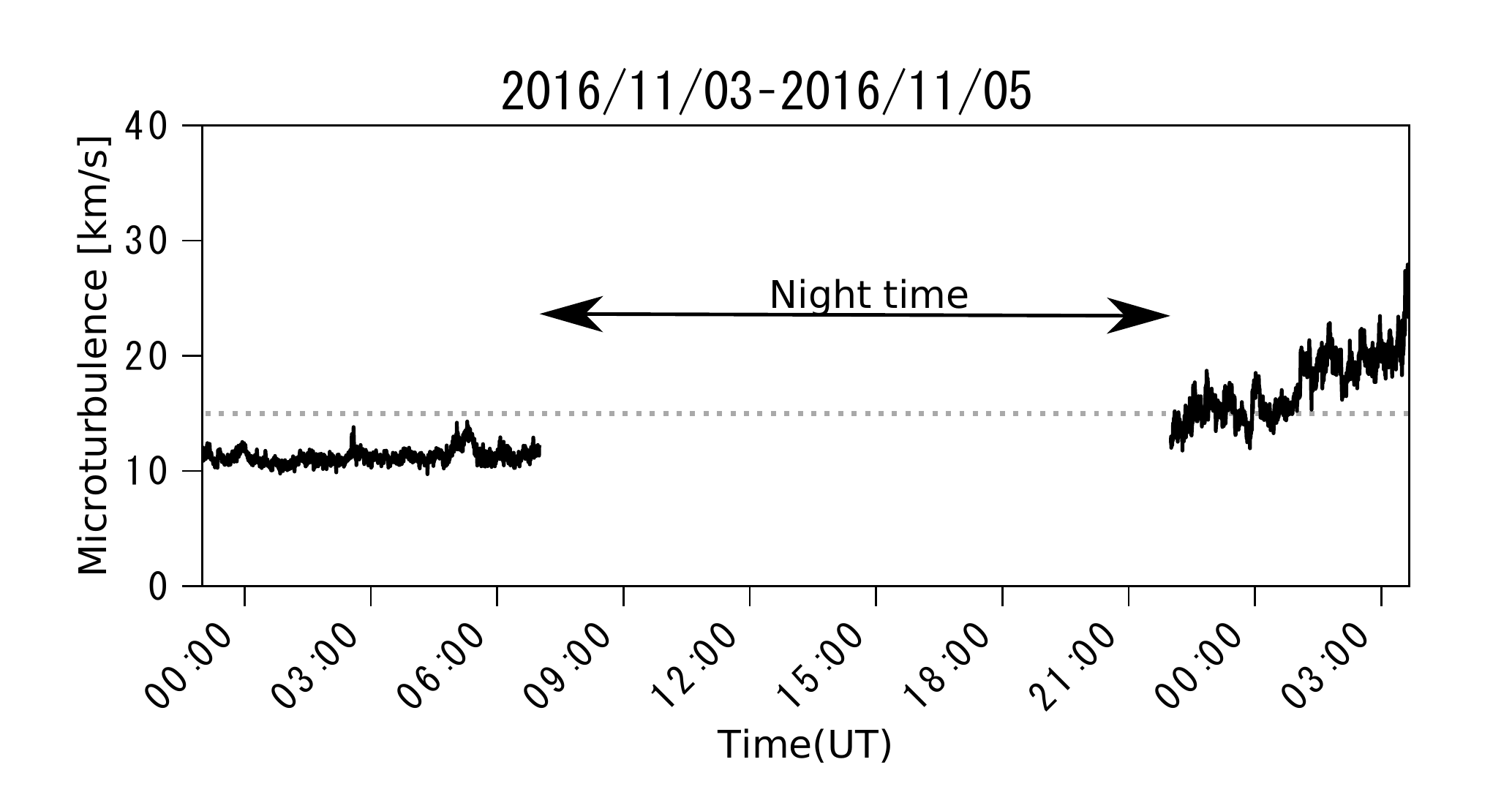}
\caption{The spatially averaged micro-turbulence of the filament from November 3 on 23:00 UT (around 29 hours prior to eruption) to November 5 on 03:40 UT (the eruption time). 
The horizontal dotted line indicates 15 km s$^{-1}$.}
\label{xi}
\end{figure}

Figure \ref{xi} shows the temporal evolution of the spatially averaged micro-turbulence of the filament derived from Equation (2).
The average was taken for the entire main body of the filament. 
(For the methodology to determine the main body of the filament, see \cite{seki17}.)
The horizontal dotted line indicates a micro-turbulence of 15 km s$^{-1}$.
Although the mean micro-turbulence had been around 12 km s$^{-1}$ until 21 hours prior to eruption, it increased to around 14 km s$^{-1}$ at 22:00 UT on November 4 (6 hours prior to eruption).
It continued increasing to around 25 km s$^{-1}$ until the total disappearance of the filament in H$\alpha$ line.

\subsection{Coronal line emission}
\begin{figure}[ht!]
\gridline{
	\fig{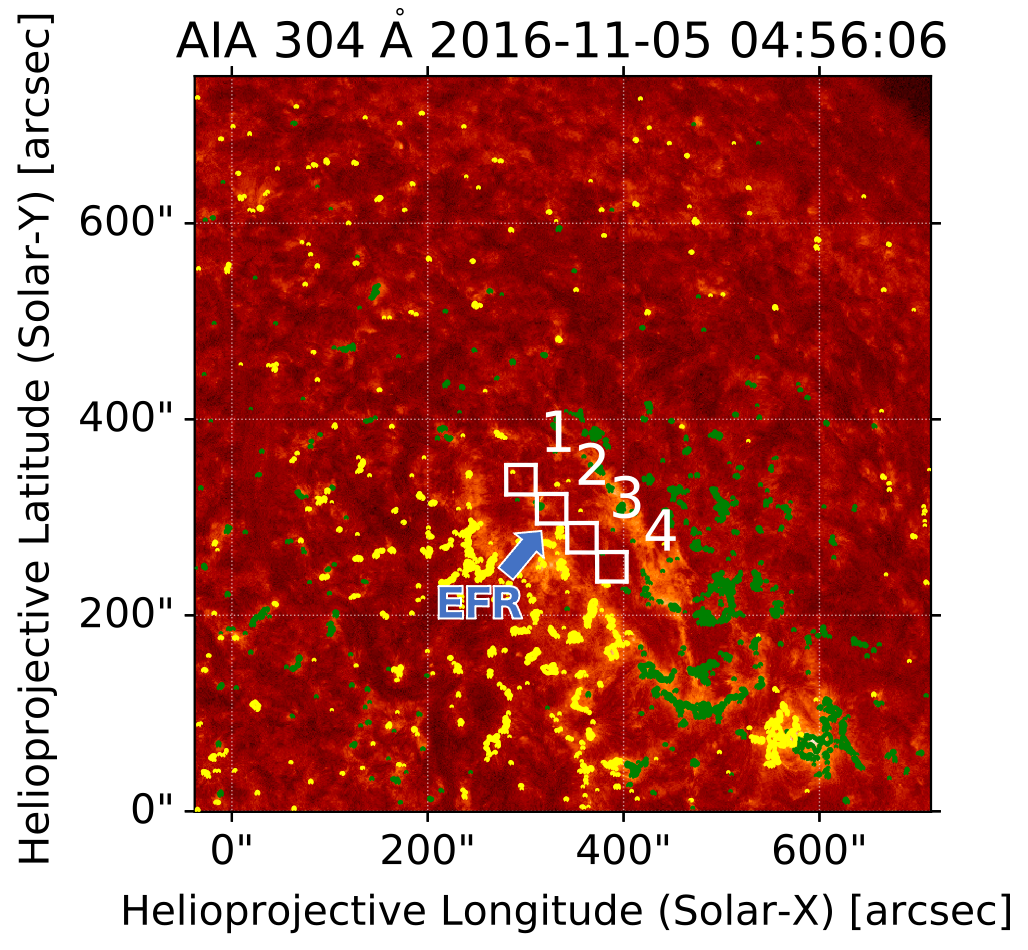}{0.6\textwidth}{(a)}
	}
\gridline{
	\fig{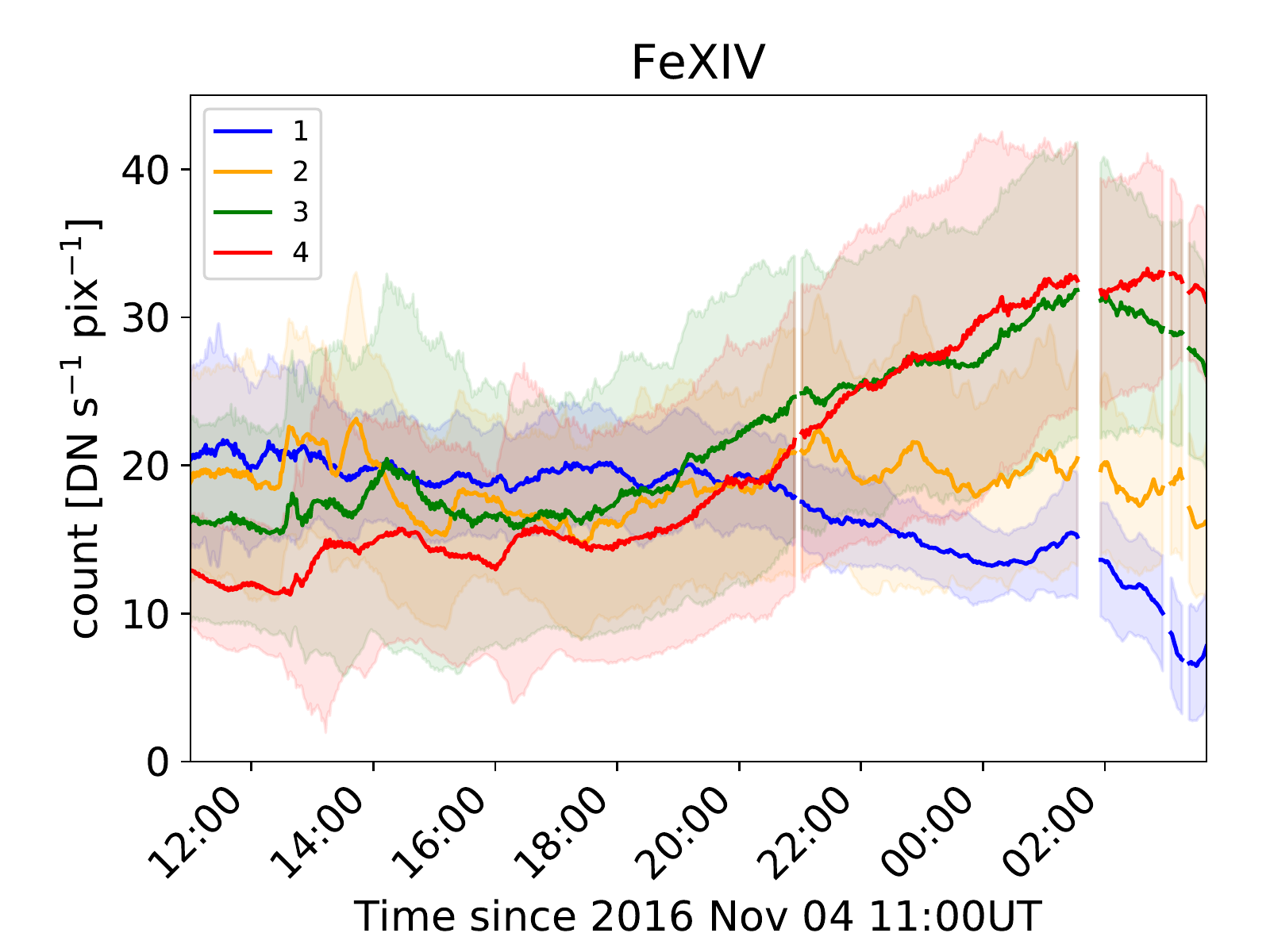}{0.7\textwidth}{(b)}
	}
\caption{
(a) The areas to calculate average counts are depicted as white rectangles on SDO/AIA 304 image.
For the coloured contours in yellow and green and EFR, see Figure \ref{efr}.
(b) Temporal evolution of spatially averaged counts per pixel in Fe XIV emission inside four different areas, 1, 2, 3, and 4.
Each number corresponds to the rectangle in the panel (a).
The effect of the solar rotation was corrected for determining the white rectangles.
The shaded area indicates $\pm$1$\sigma$ counts inside each rectangle.
}
\label{fe14}
\end{figure}

Figure \ref{fe14} shows the temporal evolution of the spatially averaged count per pixel in Fe XIV (a few 10$^{6}$ K) from a linear combination of AIA 211, 171 and 193 passbands  \citep{dz13} estimated by;
\begin{equation}
I(\textrm{Fe XIV}) = I (211 \AA) - \frac{I(171 \AA)}{17} - \frac{I(193 \AA)}{5}.
\end{equation}
White rectangles are located between the two ribbons.
These locations were in the vicinity of the stable filament seen in H$\alpha$ centre.
To improve the signal-to-noise, we temporally averaged the AIA data over 5 minutes.
We can observe rising intensities inside the rectangles 3 and 4 approximately from 18:00 UT, while those in the other regions did not exhibit such an ascent.
Note that the location of EFR in Figure \ref{efr} corresponds to the rectangle 2.

\begin{figure}[ht!]
\gridline{
	\fig{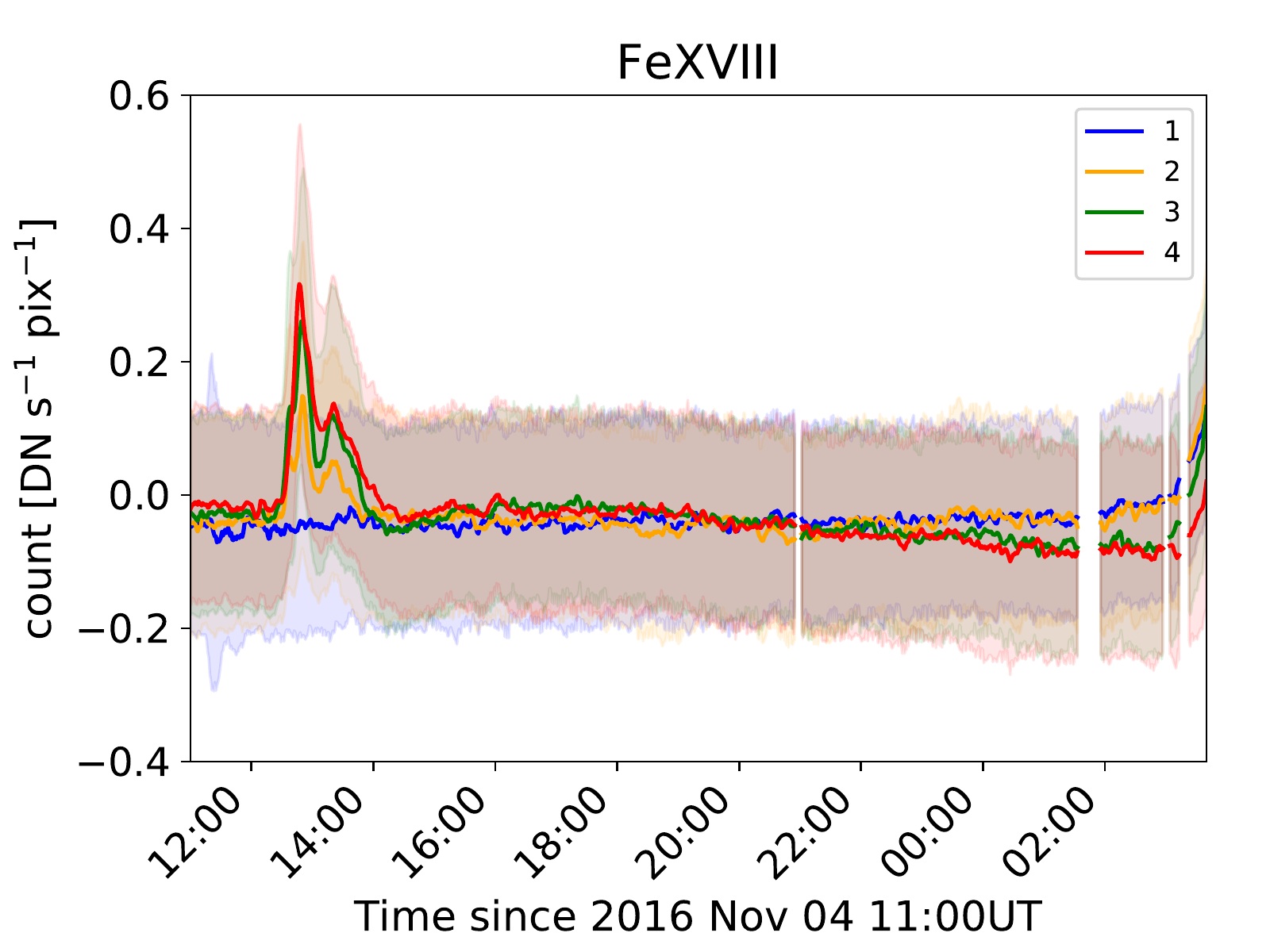}{0.7\textwidth}{}
	}
\caption{
Same plot as Figure \ref{fe14} except the emission line of Fe XVIII.
}
\label{fe18}
\end{figure}

Figure \ref{fe18} exhibits the same temporal evolution except the emission line of Fe \textrm{XVIII} ($\sim7$ MK) estimated by the equation \citep{dz13};
\begin{equation}
I(\textrm{Fe XVIII}) = I (94 \AA) - \frac{I(211 \AA)}{120} - \frac{I(171 \AA)}{450}.
\end{equation}
Inside any rectangles, the counts did not exhibit the rising feature except the sudden increase around 13:00 UT. 
This is due to a B-class flare at NOAA AR 12605 (see Figure \ref{di_height}).

\section{Discussion}\label{sec:disc}
\subsection{Turbulent motion}
In our previous work \citep{seki17}, we found that the standard deviation of the LOS velocity distribution in a filament, which indicates the amplitude of the small-scale motions, increased prior to the onset of the eruption.
As seen in Figure \ref{xi}, the mean micro-turbulence inside the filament also demonstrated an increase with time, which should reflect the turbulent motion of the plasma even within a spatial scale of one pixel.
It should be noted that
the estimate of LOS velocity assumes only one moving component of plasma blob along line of sight, whereas the micro-turbulence considers the unresolved plasma motion.
To derive a more realistic LOS velocity distribution, we need further investigation to improve the model, but that is out of the focus of this study.

Here, we suggest two possible origins for the preceding turbulent motions; one is the magnetic Rayleigh-Taylor instability, and the other is small-scale reconnections.
The small-scale vertical motions of plasma are often observed in quiescent prominences with high-resolution observations \citep{2008ApJ...676L..89B, 2011Natur.472..197B}.
\cite{2011ApJ...736L...1H} carried out three-dimensional MHD simulations to analyse how the magnetic flux rope (KS model) is stable to the magnetic Rayleigh-Taylor instability, and reconstructed the up-flows of the plasma with constant velocities. 
The dependences between initial parameters of the ideal MHD simulations and the evolution of the instability were also discussed in \cite{2012ApJ...746..120H}.
In their studies, the maximum velocity of a rising plume reached 5.9 and 2.5 km s$^{-1}$ under the conditions of plasma $\beta = $ 0.5 and 0.2, respectively.
The terminal velocity of the rising plasma plume is determined by the balance 
among the Lorentz force, the gravitational force, and the gas pressure gradient at the top of the plume.
At the beginning of the rising, the plume experiences an acceleration due to the buoyancy dominant force.
It continues until the magnetic fields are transported by the flow and sufficiently accumulated at the top of the plume to balance the gravitational force, the magnetic tension, and the magnetic and gas pressures.
As a magnetic flux rope containing a filament is reaching eruption, it is expected that the flux rope expands, and the magnetic fields become weaker.
Thus, it will take more time to realise the force balance at the top of the plume, resulting in the faster terminal velocity of the plume and the more active small-scale motions in the filament.

The other possibility is due to the small-scale reconnections below the filament.
In this study, we observed rising intensities in the coronal line in the vicinity of the filament (see Figure \ref{fe14}).
These increasing profiles could denote the continuous occurrence of small-scale reconnections below the flux rope, which could lead to destabilising it and result in a disturbance of the small-scale plasma inside the filament.
\cite{2006A&amp;A...458..965C} observed an EUV brightening feature ($\sim$10$^{6}$ K) at the footpoint of a prominence $\sim$20 min prior to the eruption.
In their study, they concluded that this suggestive brightening denoted the onset of ``tether-cutting reconnection'', in which once reconnections below a flux rope are initiated, it will ascend due to cutting off the anchoring magnetic fields, and more reconnections will be induced \citep{2001ApJ...552..833M}.
The observation of the continuous enhancement in Fe XIV could be attributable to this positive feedback process, and the evolution of this event could be explained by this scenario.
Note that during this period there was no increase in Fe XVIII line (see Figure \ref{fe18}), which indicates that there were no substantial flares causing high temperatures.

\subsection{Trigger and evolution of the eruption}
\begin{figure}[ht!]
\gridline{\fig{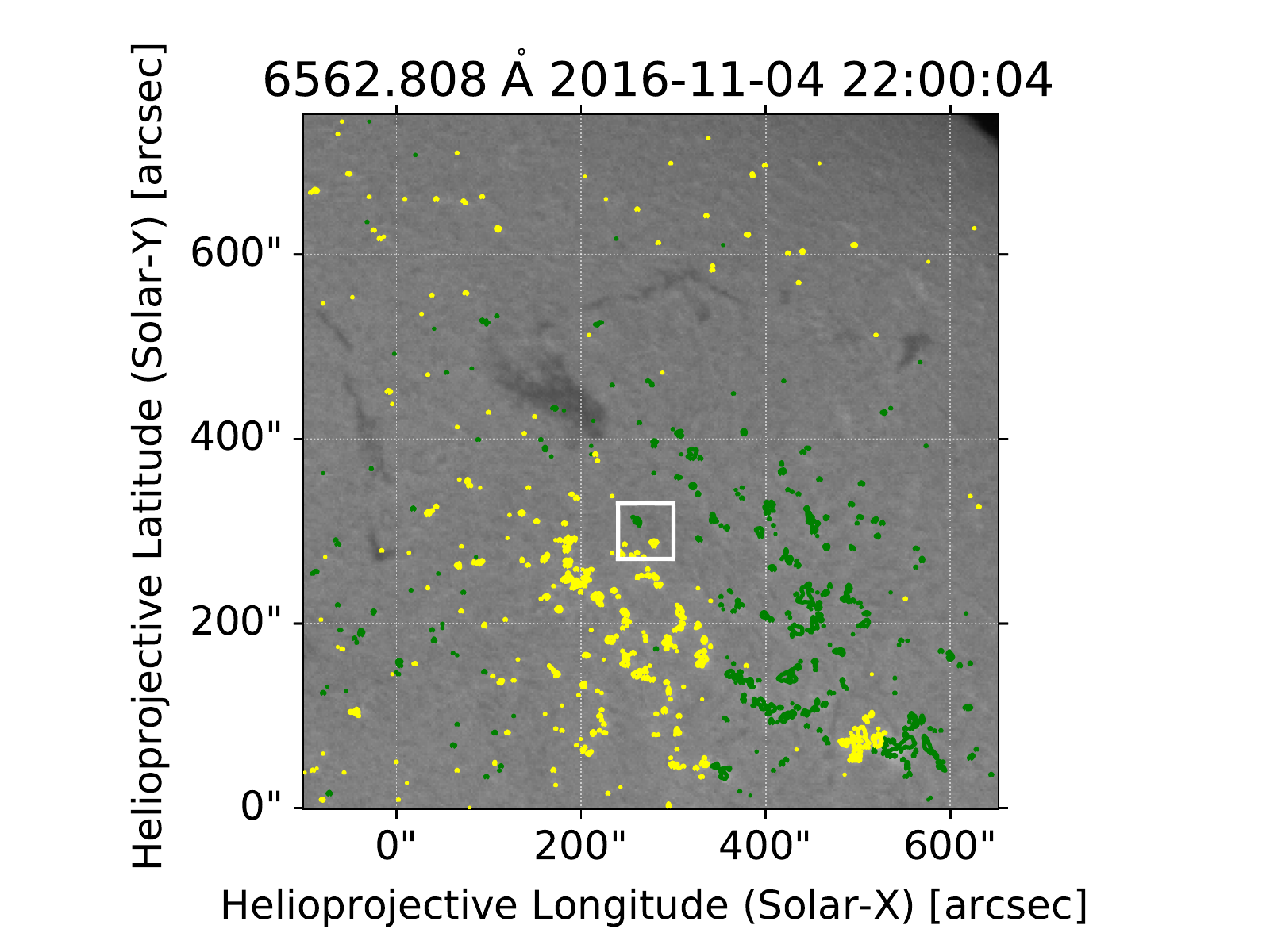}{0.5\textwidth}{}}
\gridline{\fig{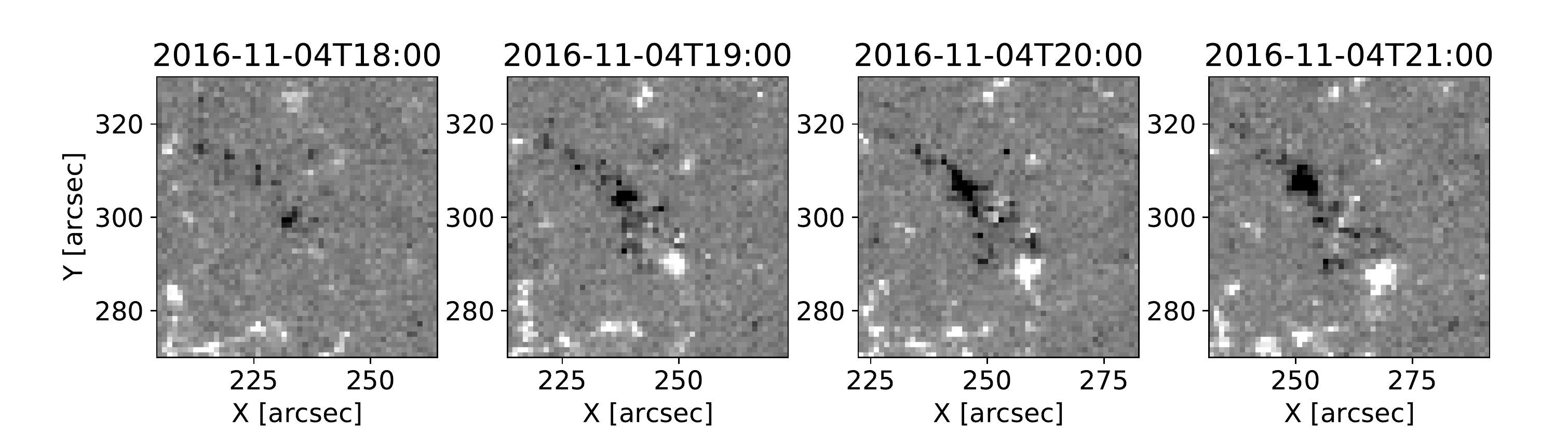}{1.0\textwidth}{}}
\caption{
Top: H$\alpha$ image taken by the SDDI superimposed by the contours of $+$100 G (yellow) and $-$100 G (green) of the HMI LOS magnetorgram. 
Bottom: The HMI LOS magnetograms inside the white rectangle in the top panel at four different times. 
To enhance visualisation, the magnetograms are shown with a scale of lower and upper limits of $\pm$ 100 G.}
\label{efr}
\end{figure}

\begin{figure}[ht!]
\plotone{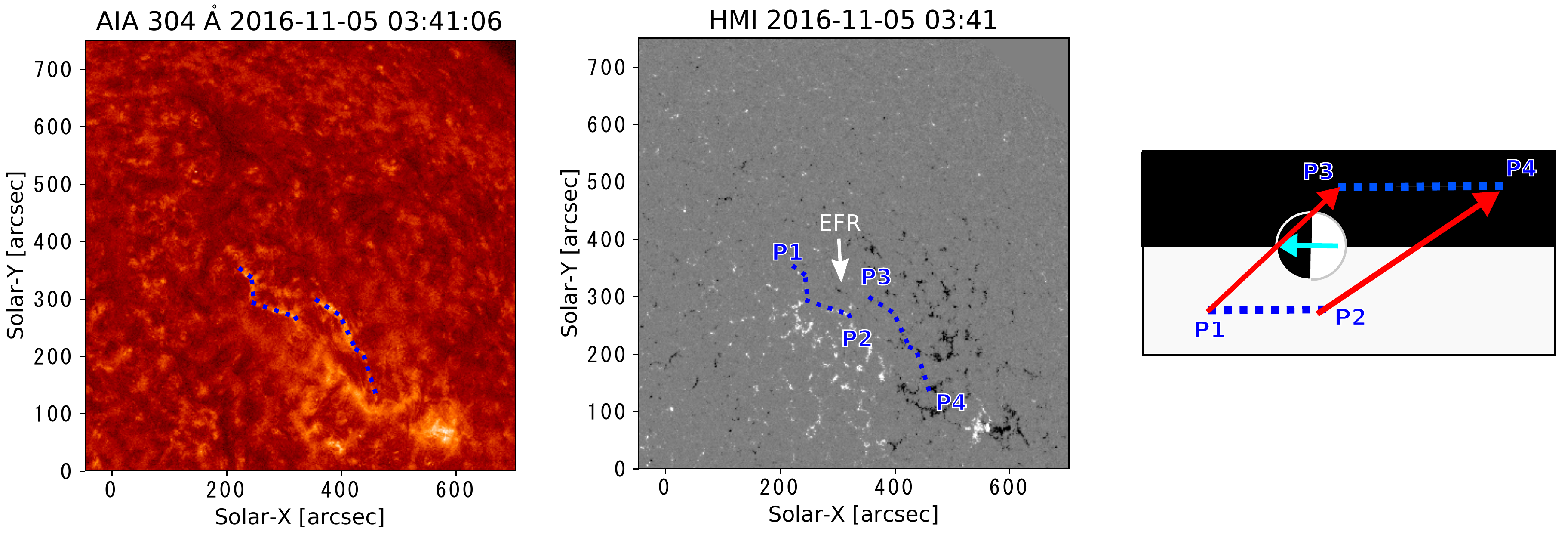}
\caption{Left and middle: Solar sub-images observed by SDO/AIA 304 and SDO/HMI at 03:41 UT on November 5.
The blue dotted line corresponds to the edge of the flare ribbon. 
White arrow indicates the emerging flux region. 
To enhance visualisation, the magnetogram is shown with a scale of lower and upper limits of $\pm$ 200 G. 
Right: The schematic diagram of top view of the large-scale and small-scale magnetic structures. 
Cyan and red arrows indicate the small-scale magnetic field and the possible large-scale overlying field, respectively.
White and black regions correspond to the positive and negative polarity regions, respectively. 
Blue dotted lines and P1--4 are simple expression of those in the middle panel.
This magnetic-filed configuration corresponds to RS-type configuration for negative shear (clockwise rotation) introduced in \cite{2012ApJ...760...31K}}
\label{shear}
\end{figure}

An emerging flux was observed around 9 hours before eruption, and this emerging flux could be the trigger of the eruption.
In fact, several studies showed that the existence of EFR plays a key role in the initiation of a filament eruption from the observational and theoretical points of view \citep{1995JGR...100.3355F, 2000ApJ...545..524C, 2012ApJ...760...31K}.
Figure \ref{efr} shows the location of a bipole in comparison with that of the filament and the snapshots of EFR observed by the HMI.
One can recognise that from 19:00 UT on November 4 the bipole evolved with time.
The two polarities were separated with time, which is characteristic of an emerging flux.
\cite{2012ApJ...760...31K} found that there are several types of small magnetic structures which should appear in the vicinity of the PIL in order to favour the onset of solar eruptions.
Especially, one of them is called reversed-shear-type (RS-type), in which small-scale magnetic field (such as emerging flux) is injected to pre-existing large-scale sheared magnetic field with a certain rotation angle with respect to large-scale potential field (see Figure 1 and 5 in \cite{2012ApJ...760...31K}).
Figure \ref{shear} shows solar sub-images observed by the AIA 304 and the HMI at 03:41 UT on November 5 superimposed by the blue dotted line indicating the edge of the two ribbons (left and middle).
We can notice that the large-scale magnetic field surrounding the filament was sheared in clockwise.
Right panel shows the schematic diagram of the top view of the large-scale and small-scale magnetic structures.
This magnetic-filed configuration corresponds to RS-type configuration for negative shear (clockwise rotation) introduced in \cite{2012ApJ...760...31K}.
Thus, we conclude that the filament eruption could be triggered by the emerging flux observed at 18:00UT.

Figure \ref{di_height} shows that the decay index nearby the filament was smaller than one, meaning that the flux rope was stable against the torus instability .
It should be noted that in \cite{2006PhRvL..96y5002K}, they assumed the shape of flux rope as a ring. 
On the other hand, \cite{ishiguro17} found that the ideal MHD instability can be initiated even in the torus-stable condition, i.e. decay index is less than one, if the magnetic loop has a double-arc-shape configuration.
They called this critical condition for the eruption under a certain geometry as Double arc instability (DAI).
The DAI-favoured magnetic configuration can be produced in the ``tether-cutting reconnection'' scenario \citep{2001ApJ...552..833M}.
Additionally, this scenario agrees with the result of \cite{2012ApJ...760...31K}.
The evolution of this event could be explained by this scenario (see the previous section), and the magnetic-field configuration could be the DAI-favoured one.

\subsection{Space weather application}
Quiescent filament eruptions occasionally drive large CMEs and cause severe geomagnetic disturbances.
Thus, in terms of space weather prediction, it is also of a great importance to predict filament eruptions.
\cite{1996JGR...10113497M} reported a large polar crown filament eruption on 1994 April 14. 
Although this event was not associated with any significant flares, we experienced a severe geomagnetic disturbance (Dst $\sim$ -200 nT) within a few days of the eruption.
\cite{isobe19} investigated the records of aurora display in the middle magnetic latitudes (China and Japan) during the Maunder minimum in 1653, which indicates the presence of a great geomagnetic disturbance although the solar activity at that time must have been very quiet.
With a simple theoretical discussion, they concluded that this geomagnetic storm was extremely intense (Dst $<$ $-$300 nT) and can be driven by a quiescent filament eruption.

Our result implies that it would have been impossible to predict the filament eruption only from the photospheric magnetic fields and the extrapolated potential fields for this event because the photospheric magnetic fields scarcely changed over the past several days of the eruption.
From Figure \ref{di_height}, we cannot recognise any significant changes in the global magnetic-field configuration which lead to the onset of  the eruption.
Moreover, the region to be considered around the filament is a global quiet one, and we can hardly expect to obtain precise vector-magnetic-field data. 
That is why it is difficult to extrapolate more realistic magnetic fields such as non-linear force free fields for this event.
This illustrates that there is a case in which it is difficult to predict and monitor when a filament will erupt only from the global magnetic-field configuration. 
Thus, we suggests that the internal turbulent motion in a filament can also provide useful clues to predict filament eruptions \citep{seki19}.

\acknowledgements{
D.S. thanks an anonymous referee for significantly improving our manuscript.
D.S. thanks Dr. T. Kaneko and Dr. A. Hillier for the discussion of the origin of the turbulence.
D.S. also thanks GSAIS Empirical Research Group (GERG) for the discussion to enrich the methodology.
We express our sincere gratitude to the staff of the Hida Observatory for development and maintenance of the instrument and daily observation.
This work was supported by JSPS KAKENHI Grant numbers JP15H05814 (Project for Solar-Terrestrial Environment Prediction, PSTEP), JP16H03955, and JP18J23112. 
D. S. is supported by Research Fellowships of Japan Society for the Promotion of Science for Young Scientists.
T. S. is supported by JSPS KAKENHI Grant Number JP18J12677.
GDZ acknowledges support from STFC (UK) via the consolidated grants to the atomic astrophysics group (AAG) at DAMTP, University of Cambridge (ST/P000665/1 and ST/
T000481/1).
This research utilised SunPy, an open-source and free community-developed solar data analysis package written in Python \citep{sunpy}.
}
\software{
Potential Field Source Surface Solver \citep{yeates17},
SunPy \citep{sunpy}
}




\begin{thebibliography}{}
\bibitem[Antiochos et al.(1999)]{1999ApJ...510..485A}Antiochos, S. K., DeVore, C. R., \& Klimchuk, J. A. 1999, ApJ, 510, 485
\bibitem[Bateman (1978)]{bate78}Bateman, G. 1978, MHD Instabilities (Cambridge, MA: MIT Press), p.85
\bibitem[Beckers (1964)]{1964beckers}Beckers, J. 1964, Ph.D. Thesis, University of Utrecht
\bibitem[Berger et al. (2008)]{2008ApJ...676L..89B}Berger, T. E., Shine, R. A., Slater, G. L., et al. 2008, \apjl, 676, L89
\bibitem[Berger et al. (2011)]{2011Natur.472..197B}Berger, T., Testa, P., Hillier, A., et al. 2011, \nat, 472, 197
\bibitem[Brueckner et al. (1995)]{lasco}Brueckner, G. E., Howard, R. A., Koomen, M. J., et al. 1995, \solphys, 162, 357
\bibitem[Cabezas et al. (2017)]{2017ApJ...836...33C}Cabezas, D. P., Mart\'inez, L. M., Buleje, Y. J., et al. 2017, \apj, 836, 33
\bibitem[Cane \& Richardson (2003)]{cr03}Cane, H. V. \& Richardson, I. G. 2003, \jgr, 108, 1156
\bibitem[Chen \& Shibata(2000)]{2000ApJ...545..524C}Chen, P. F., \& Shibata, K. 2000, \apj, 545, 524
\bibitem[Chifor et al.(2006)]{2006A&amp;A...458..965C}Chifor, C., Mason, H. E., Tripathi, D., et al. 2006, A\&A, 458, 965
\bibitem[Cliver et al. (2009)]{cliver09}Cliver, E. W., Balasubramaniam, K. S., Nitta, N. V., et al. 2009, \jgr, 114, A00A20
\bibitem[Del Zanna (2013)]{dz13}Del Zanna, G., 2013, \aap, 558, A73
\bibitem[Feynman \& Martin (1995)]{1995JGR...100.3355F}Feynman, J., \& Martin, S. F. 1995, \jgr, 100, 3355
\bibitem[Filippov \& Den (2001)]{2001JGR...10625177F}Filippov, B. P., \& Den, O. G. 2001, \jgr, 106, 25177
\bibitem[Filippov (2013)]{filippov13}Filippov, B. 2013, \apj, 773, 10
\bibitem[Gilbert et al. (2001)]{gilbert01} Gilbert, H. R., Holzer, T. E., Low, B. C., et al. 2001, \apj, 549, 1221
\bibitem[Gopalswamy et al. (2000)]{gopal00}Gopalswamy, N., Lara, A., Lepping, R. P., et al. 2000, \grl, 27, 145
\bibitem[Gosain et al.(2009)]{2009SoPh..259...13G}Gosain, S., Schmieder, B., Venkatakrishnan, P., Chandra, R., \& Artzner, G. 2009, SoPh, 259, 13
\bibitem[Harra et al.(2001)]{2001ApJ...549L.245H}Harra, L. K., Matthews, S. A., \& Culhane, J. L. 2001, \apjl, 549, L245
\bibitem[Harra et al.(2009)]{2009ApJ...691L..99H}Harra, L. K., Williams, D. R., Wallace, A. J., et al. 2009, \apjl, 691, L99
\bibitem[Harvey et al. (2011)]{gong}Harvey, J.W., Bolding, J., Clark, R., et al. 2011, \baas, 43, 17
\bibitem[Hillier et al. (2011)]{2011ApJ...736L...1H}Hillier, A., Isobe, H., Shibata, K., et al. 2011, \apjl, 736, L1
\bibitem[Hillier et al. (2012)]{2012ApJ...746..120H}Hillier, A., Berger, T., \& Isobe, H. 2012, \apj, 746, 120
\bibitem[Hillier \& Arregui (2019)]{ah19}Hillier, A. \& Arregui, I., 2019, \apj, in press (arXiv:1909.11351)
\bibitem[Ichimoto et al. (2017)]{2017SoPh..292...63I}Ichimoto, K., Ishii, T. T., Otsuji, K., et al. 2017, \solphys, 292, 63
\bibitem[Ishiguro \& Kusano (2017)]{ishiguro17}Ishiguro, N., \& Kusano, K. 2017, \apj, 843, 101
\bibitem[Isobe \& Tripathi(2006)]{isobe06}Isobe, H., \& Tripathi, D. 2006, \aap, 449, L17
\bibitem[Isobe et al. (2007)]{2007SoPh..246...89I}Isobe, H., Tripathi, D., Asai, A., et al. 2007, \solphys, 246, 89
\bibitem[Isobe et al. (2019)]{isobe19}Isobe, H., Ebihara, Y., Kawamura, A. D., et al. 2019, \apj, in press (arXiv:1903.08466)
\bibitem[Joselyn \& McIntosh (1981)]{joselyn81}Joselyn, J. A. \& McIntosh, P. S. 1981, \jgr, 86, 4555
\bibitem[Joshi et al. (2014a)]{joshi14a}Joshi, N. C., Srivastava, A. K., Filippov, B., et al. 2014, \apj, 787, 11
\bibitem[Joshi et al. (2014b)]{joshi14b}Joshi, N. C., Magara, T., \& Inoue, S. 2014, \apj, 795, 4
\bibitem[Kasper (2002)]{02kasper}Kasper, J. C., 2002, Ph.D. Thesis, Massachusetts Institute of Technology
\bibitem[King \& Papitashvili (2005)]{omni} King, J. H., \& Papitashvili, N. E. 2005, \jgr, 110, A02104
\bibitem[Kippenhahn \& Shul\"uter (1957)]{ksmodel} Kippenhahn, R., \& Schl\"uter, A. 1957, \zap, 43, 36
\bibitem[Kliem \& T\"or\"ok (2006)]{2006PhRvL..96y5002K}Kliem, B., \& T\"or\"ok, T. 2006, Physical Review Letters, 96, 255002
\bibitem[Kuperus \& Raadu (1974)]{krmodel}Kuperus, M., \& Raadu, M. A. 1974, \aap, 31, 189
\bibitem[Kusano et al.(2012)]{2012ApJ...760...31K}Kusano, K., Bamba, Y., Yamamoto, T. T., et al. 2012, \apj, 760, 31
\bibitem[Lemen et al. (2012)]{2012SoPh..275...17L}Lemen, J. R., Title, A. M., Akin, D. J., et al. 2012, \solphys, 275, 17
\bibitem[Li et al. (2016)]{li16}Li, H., Liu, Y., Elmhamdi, A. et al. 2016, \apj, 830, 132
\bibitem[Magara \& Tsuneta(2008)]{2008PASJ...60.1181M}Magara, T., \& Tsuneta, S. 2008, PASJ, 60, 1181
\bibitem[McAllister et al. (1996)]{1996JGR...10113497M}McAllister, A. H., Dryer, M., McIntosh, P., et al. 1996, \jgr, 101, 13497
\bibitem[McCauley et al. (2015)]{2015SoPh..290.1703M}McCauley, P. I., Su, Y. N., Schanche, N., et al. 2015, \solphys, 290, 1703
\bibitem[Moore et al.(2001)]{2001ApJ...552..833M}Moore, R. L., Sterling, A. C., Hudson, H. S., et al. 2001, \apj, 552, 833
\bibitem[Morimoto \& Kurokawa (2003a)]{2003PASJ...55..503M}Morimoto, T., \& Kurokawa, H. 2003, \pasj, 55, 503 
\bibitem[Morimoto \& Kurokawa (2003b)]{2003PASJ...55.1141M}Morimoto, T., \& Kurokawa, H. 2003b, \pasj, 55, 1141
\bibitem[Morimoto et al. (2010)]{2010PASJ...62..939M}Morimoto, T., Kurokawa, H., Shibata, K., et al. 2010, \pasj, 62, 939
\bibitem[Nagashima et al. (2007)]{2007ApJ...668..533N}Nagashima, K., Isobe, H., Yokoyama, T., et al. 2007, \apj, 668, 533
\bibitem[National Research Council (2008)]{nrc08}National Research Council 2008, Severe Space Weather Events: Understanding Societal and Economic Impacts: A Workshop Report, (Washington, D.C: The National Academies Press)
\bibitem[Nose et al. (2015)]{dstindex}Nose, M., Sugiura, M., Kamei, T., et al. 2015,  Geomagnetic Dst index, World Data Center for Geomagnetism, Kyoto, doi:10.17593/14515-74000
\bibitem[Ofman et al. (1998)]{ofman98}Ofman, L., Kucera, T. A., Mouradian, Z., \& Poland, A. I. 1998, \solphys, 183, 97
\bibitem[Ohyama \& Shibata (1997)]{ohshi97}Ohyama, M. \& Shibata, K. 1997, \pasj, 49, 249
\bibitem[Parenti (2014)]{parenti14}Parenti, S. 2014, LRSP, 11, 1
\bibitem[Sakaue et al. (2018)]{sakaue18}Sakaue, T., Tei, A., Asai, A., et al. 2018, \pasj, 70, 99
\bibitem[Scherrer et al. (2012)]{scherrer12}Scherrer, P.H., Schou, J., Bush, R.I., et al. 2012, \solphys, 275, 207
\bibitem[Sterling \& Moore (2004)]{slingmo04}Sterling, A. C. \& Moore, R. L. 2004, \apj, 602, 1024
\bibitem[Seki et al. (2017)]{seki17}Seki, D., Otsuji, K., Isobe, H., et al. 2017, \apj, 843, L24
\bibitem[Seki et al. (2018)]{seki18}Seki, D., UeNo, S., Isobe, H., et al. 2018, SunGe, 13, 157
\bibitem[Seki et al. (2019)]{seki19}Seki, D., Otsuji, K., Isobe, H., et al. 2019, \pasj, doi:10.1093/pasj/psz031
\bibitem[Shibata \& Magara (2011)]{2011LRSP....8....6S}Shibata, K., \& Magara, T. 2011, LRSP, 8, 6
\bibitem[Smith \& Ramsey (1964)]{SmithRamsey64}Smith, S. F. \& Ramsey, H. E. \ 1964, \zap, 60, 1
\bibitem[Sterling \& Moore (2004)]{2004ApJ...602.1024S} Sterling, A. C., \& Moore, R. L. 2004, \apj, 602, 1024
\bibitem[Sterling et al. (2011a)]{sterling11}Sterling, A. C., Moore, R. L., \& Freeland, S. L. 2011, \apjl, 731, L3
\bibitem[Sterling et al. (2011b)]{2011ApJ...743...63S}Sterling, A. C., Moore, R. L., \& Harra, L. K. 2011, \apj, 743, 63
\bibitem[The SunPy Community et al. (2015)]{sunpy}The SunPy Community, Mumford, S. J., Christe, S., et al. 2015, Computational Science {\&} Discovery, 8, 014009
\bibitem[Tandberg-Hanssen (1995)]{tandberg1995nature}Tandberg-Hanssen, E. 1995, The Nature of Solar Prominences, Astrophysics and Space Science Library (Springer)
\bibitem[UeNo et al.(2004)]{2004SPIE.5492..958U}UeNo, S., Nagata, S., Kitai, R., et al. 2004, in Proc. SPIE, Vol. 5492, Ground-based Instrumentation for Astronomy, ed. A. F. M. Moorwood \& M. Iye, 958--969
\bibitem[UeNo et al. (2007)]{ueno07}UeNo, S., Shibata, K., Kimura, G., et al. 2007, BASI, 35, 697
\bibitem[van Ballegooijen et al. (2000)]{vB00}van Ballegooijen, A. A., Priest, E. R., \& Mackay, D. H. 2000, \apj, 539, 983
\bibitem[Yeates (2017)]{yeates17}Yeates, A. R. 2017, pfss.py, https://github.com/antyeates1983/pfss
\bibitem[Zuccarello et al. (2016)]{zucc16}Zuccarello, F. P., Aulanier, G., \& Gilchrist, S. A. 2016, \apjl, 821, L23
\end{thebibliography}
\end{document}